\documentclass{article}
\usepackage{orcidlink}

\usepackage{microtype}
\usepackage{graphicx}
\usepackage{subcaption}
\usepackage{booktabs} % for professional tables

\usepackage{hyperref}

\usepackage[accepted]{icml2026}

\usepackage{amsmath}
\usepackage{amssymb}
\usepackage{mathtools}
\usepackage{amsthm}

\usepackage[capitalize,noabbrev]{cleveref}

\theoremstyle{plain}

\theoremstyle{definition}

\theoremstyle{remark}

\usepackage[textsize=tiny]{todonotes}

\icmltitlerunning{Which Anatomy Matters Under Limited Labels? A Data-Efficient Anatomy-Aware Benchmark for Cardiac Pathology Prediction}

\begin{document}

\twocolumn[
\icmltitle{Which Anatomy Matters Under Limited Labels? A Data-Efficient Anatomy-Aware Benchmark for Cardiac Pathology Prediction}
  %\icmltitle{Which Anatomy Matters Under Limited Labels? An Anatomy-Aware Benchmark for Low-Data Cardiac Pathology Prediction}

  % It is OKAY to include author information, even for blind submissions: the
  % style file will automatically remove it for you unless you've provided
  % the [accepted] option to the icml2026 package.

  % List of affiliations: The first argument should be a (short) identifier you
  % will use later to specify author affiliations Academic affiliations
  % should list Department, University, City, Region, Country Industry
  % affiliations should list Company, City, Region, Country

  % You can specify symbols, otherwise they are numbered in order. Ideally, you
  % should not use this facility. Affiliations will be numbered in order of
  % appearance and this is the preferred way.
  \icmlsetsymbol{equal}{*}

  \begin{icmlauthorlist}
    \icmlauthor{Himanshu Singh \orcidlink{0000-0001-8816-1340}}{equal,yyy}
    %\icmlauthor{Firstname2 Lastname2}{equal,yyy,comp}
    %\icmlauthor{Firstname3 Lastname3}{comp}
    %\icmlauthor{Firstname4 Lastname4}{sch}
    %\icmlauthor{Firstname5 Lastname5}{yyy}
    %\icmlauthor{Firstname6 Lastname6}{sch,yyy,comp}
    %\icmlauthor{Firstname7 Lastname7}{comp}
    %\icmlauthor{}{sch}
    %\icmlauthor{Firstname8 Lastname8}{sch}
    %\icmlauthor{Firstname8 Lastname8}{yyy,comp}
    %\icmlauthor{}{sch}
    %\icmlauthor{}{sch}
  \end{icmlauthorlist}

  \icmlaffiliation{yyy}{Computing \& Artificial Intelligence Division, Los Alamos National Laboratory, Los Alamos, NM 87545, USA}
  %\icmlaffiliation{comp}{Company Name, Location, Country}
  %\icmlaffiliation{sch}{School of ZZZ, Institute of WWW, Location, Country}

  \icmlcorrespondingauthor{Himanshu Singh (\textcolor{blue}{\tt{https://himanshuvnm.github.io/}})}{singh\_h@lanl.gov}
  %\icmlcorrespondingauthor{Firstname2 Lastname2}{first2.last2@www.uk}

  % You may provide any keywords that you find helpful for describing your
  % paper; these are used to populate the "keywords" metadata in the PDF but
  % will not be shown in the document
  \icmlkeywords{Machine Learning, ICML}

  \vskip 0.3in
]

% this must go after the closing bracket ] following \twocolumn[ ...

% This command actually creates the footnote in the first column listing the
% affiliations and the copyright notice. The command takes one argument, which
% is text to display at the start of the footnote. The \icmlEqualContribution
% command is standard text for equal contribution. Remove it (just {}) if you
% do not need this facility.

% Use ONE of the following lines. DO NOT remove the command.
% If you have no special notice, KEEP empty braces:
\printAffiliationsAndNotice{}  % no special notice (required even if empty)
% Or, if applicable, use the standard equal contribution text:
% \printAffiliationsAndNotice{\icmlEqualContribution}

\begin{abstract}
Numerous medical imaging problems must be solved under limited labels and constrained compute, yet it remains unclear whether performance gains are driven mainly by more expressive models or by better representation of clinically meaningful anatomy. We study this question through a low-data anatomy-aware benchmark for 5-class cardiac pathology prediction on the public ACDC MRI dataset. Using segmentation-derived patient descriptors from the right ventricle, myocardium, and left ventricle, we compare anatomy-specific and multi-structure representations across linear, kernel, and tree-based classifiers. We find that under limited label settings, representation dominates complexity. %myocardium is the strongest single-structure source of signal, the combined multi-structure representation performs best overall, and nonlinear classifiers provide only modest gains once the representation is fixed. Simple handcrafted inter-phase dynamic summaries do not improve over static anatomy-aware features. 
These results suggest that in resource-constrained healthcare settings, identifying and representing the most informative anatomy may matter more than the increasing complexity of the model alone.
\end{abstract}

\section{Introduction}

\begin{figure}[t]
    \centering
    \frame{\includegraphics[width=\linewidth]{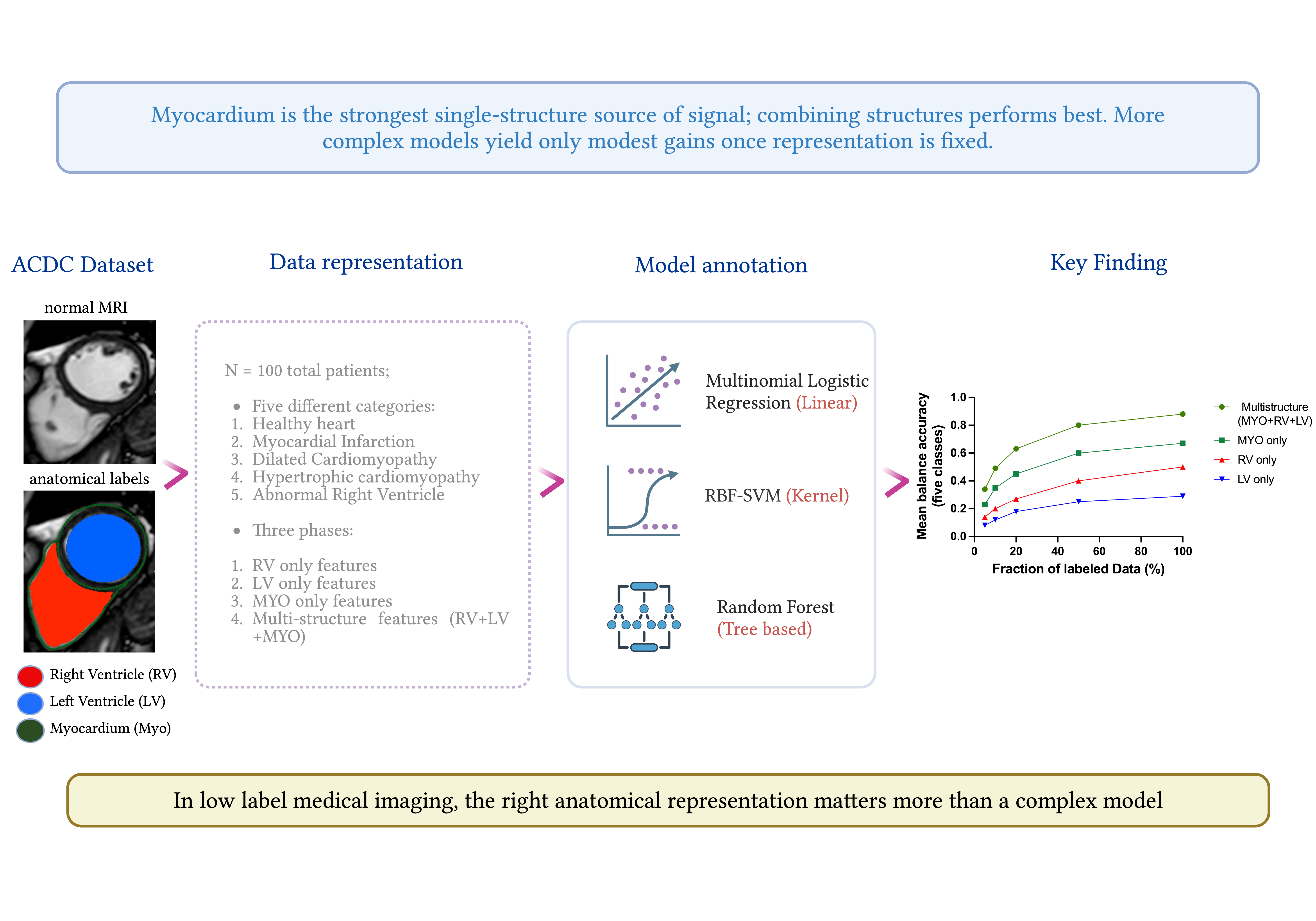}}
    \caption{\textbf{Representation before complexity.}
Our benchmark asks whether, under limited labels, selecting the right anatomical representation matters more than increasing model complexity. We illustrate this by decomposing a representative short-axis cardiac MR image into RV-only, MYO-only, LV-only, and ALL-structures views, which form the basis of the anatomy ablation study. Background cardiac MR image adapted from \citet{adomat2026cardiac}.}
%\caption{\textbf{A schematic workflow cartoon highlighting \emph{representation dictates the model complexity} in the low-label ACDC MRI dataset.} 
%Primary anatomical feature configurations used in our benchmark. The figure is constructed from a representative ACDC subject by selecting one labeled cardiac phase and displaying the slice with the largest total annotated area. We then visualize the same underlying MRI slice under four structure-selection settings: RV-only, MYO-only, LV-only, and ALL-structures (RV+MYO+LV). These configurations isolate structure-specific signal and motivate the anatomy ablation study.
%}
\label{fig:fig_hook}
\end{figure}
In the low-label medical imaging regime \citep{jin2026learning,zhou2021review}, researchers often attempt to improve performance using more complex models. However, it remains unclear whether the real bottleneck is the complexity of the model \citep{varoquaux2022machine} or how the clinical structure is represented. In particular, in cardiac imaging \citep{buja2022cardiovascular,counseller2023recent,flachskampf2015cardiac}, this question is especially relevant because the pathology is expressed through anatomically meaningful structure rather than arbitrary input variation. 

On the other hand, in AI for medical imaging, practical bottlenecks often lie not only in model design but also in data preparation, annotation, and deployment infrastructure \citep{willemink2020preparing}. These constraints are especially acute in resource-constrained healthcare settings, where the limited radiology infrastructure can hinder the adoption of compute-intensive AI pipelines \citep{yousef2024artificial}. %\subsection{Contributions}

In the present work, inspired by previous benchmark-driven progress in medical imaging \citep{bernard2018deep,blagec2023benchmark}, we address this research direction through a reproducible low-data benchmark built on Automated Cardiac Diagnosis Challenge (ACDC) MRI dataset \citep{bernard2018deep}. The intuition behind our benchmark is illustrated in \autoref{fig:fig_hook}, where a representative cardiac MR image is decomposed into structure-specific views to address which anatomy carries the dominant predictive signal.

This question is also well motivated by prior cardiac MRI diagnosis pipelines built on segmentation-derived representations. In particular, Isensee et al.~\citep{isensee2017automatic} and Khened et al.~\citep{khened2019fully} combined segmentation outputs with clinically inspired handcrafted features for automatic disease assessment on ACDC, while Zheng et al.~\citep{zheng2019explainable} showed that explainable cardiac pathology classification can be achieved by combining shape-related features with motion characterization. These works demonstrate that anatomically grounded descriptors can support interpretable pathology prediction; our focus is to understand, under limited labels, which anatomical structures carry the dominant predictive signal and how much that matters relative to classifier complexity.

\paragraph{Our Contributions.}

We construct an anatomy-aware benchmark for low-data cardiac pathology prediction using ACDC segmentation masks. We then show that the myocardial morphology is the strongest single-structure source of predictive signal, while multi-structure anatomical representation yields the best overall performance. %We also demonstrate that the representation choice matters more than classifier complexity (linear, kernel, and tree-based models) achieving similar performance once the anatomical representation is fixed. 
Finally, we show that simple handcrafted inter-phase delta features do not improve over static multi-structure descriptors, and we validate the benchmark using label-shuffle controls, confusion analysis, and patient-level visualization.

\section{Anatomy-Aware Benchmark Setup \& Models}

\subsection{Dataset and Task}
%We use the public ACDC dataset and consider its balanced 5-class pathology setting: $\{\texttt{DCM},\ \texttt{HCM},\ \texttt{MINF},\ \texttt{NOR},\ \texttt{RV}\}.$ Each class contains 20 patients, yielding a total of 100 subjects. 
To begin with, we consider the balanced 5-class pathology setting on ACDC, which are dilated cardiomyopathy (DCM), hypertrophic cardiomyopathy (HCM), myocardial infarction (MINF), normal subjects (NOR), and abnormal right ventricle (RV). In this dataset, each class contains 20 patients, giving a total of 100 subjects. For brevity, we use the abbreviations DCM, HCM, MINF, NOR, and RV throughout the remainder of the paper. For each patient, we use the annotated segmentation masks provided at the labeled cardiac phases and build patient-level features from the three principal anatomical structures: {RV}, {MYO}, {LV}. 
\paragraph{Task.}
Let \(\{(x_i, y_i)\}_{i=1}^N\) denote the patient-level dataset, where \(y_i \in \{1,\dots,K\}\) is the pathology label, and \(x_i\) denotes the cardiac MRI study for patient \(i\);  \(N\) is the number of patients and \(K\) denotes the number of classes. From each study, we construct anatomy-aware feature representations
\[
\phi_r(x_i) \in \mathbb{R}^{d_r}, \qquad r \in \mathcal{R},
\]
where \(\mathcal{R} \coloneqq \{\text{RV-only},~\text{MYO-only},~\text{LV-only},~ \text{ALL}\}\). For each representation \(r\), we train classifiers \(f_{\theta}^{(r)} \in \mathcal{F}\), where \(\mathcal{F}\) includes linear, kernel, and tree-based models, to predict
\[
\hat y_i = f_{\theta}^{(r)}(\phi_r(x_i)).
\]
Our goal is not only to maximize predictive performance, but also to identify which anatomical representation \(r\) contributes the strongest signal under limited labels and whether the variation across representations is larger than the variation across classifier families.
%Our \emph{goal} is to predict multiclass pathology under limited labels. This setup allows us to study not only predictive performance but also which anatomical structures are most informative and whether more complex classifiers are necessary once the representation is fixed.

\subsection{Patient-Level Anatomical Features}

\begin{figure}[h]
    \centering
    \frame{\includegraphics[width=\linewidth]{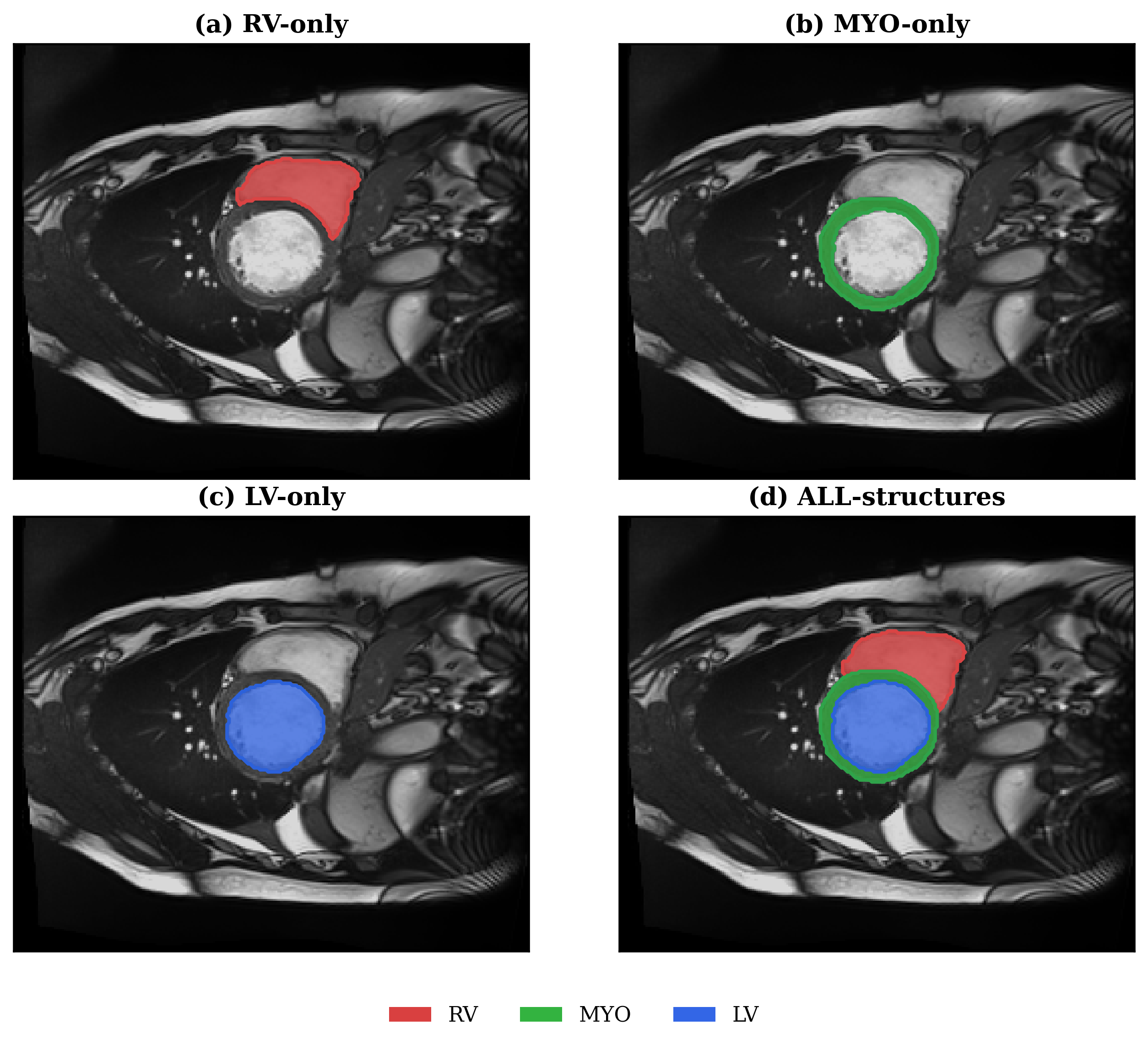}}
    \caption{\textbf{A representative visual for the primary anatomical feature configurations used in the benchmark.} %For reference, {\tt{PATIENT\_ID = "patient001"}}.
    %A representative ACDC short-axis cardiac MR slice is shown under four structure-selection settings: RV-only, MYO-only, LV-only, and ALL-structures (RV+MYO+LV). These configurations isolate structure-specific signal and form the basis of the anatomy ablation study.
    }
%\caption{%\textbf{Primary anatomical feature configurations used in our benchmark from a ACDC dataset.} %The figure is constructed from a representative ACDC subject by selecting one labeled cardiac phase and displaying the slice with the largest total annotated area. We then visualize the four feature configurations used throughout the paper on the same underlying MRI slice: RV-only, MYO-only, LV-only, and ALL-structures (RV+MYO+LV). These configurations isolate structure-specific signal and provide the basis for the anatomy ablation study.}
\label{fig:fig_feature_configurations}
\end{figure}

In order to isolate the structure-specific signal, we define four primary feature configurations: RV-only, MYO-only, LV-only, and ALL-structures, where the latter concatenates descriptors from all three anatomical compartments; refer \autoref{fig:fig_feature_configurations} for a representative visual.

\autoref{fig:fig_feature_configurations} is constructed from a representative ACDC subject by selecting a labeled cardiac phase and displaying the slice with the largest total annotated area. The same underlying MRI slice is then shown in four structure-selection settings: RV-only, MYO-only, LV-only, and ALL-structures. For each labeled frame and each anatomical structure, we extract simple shape descriptors from the binary segmentation mask, including area, area fraction, aspect ratio, principal-axis statistics, elongation, compactness, circularity, extent, and radial distance summaries. Slice-wise features are aggregated to the patient level by mean and standard deviation across the slices, together with the number of slices containing the structure.

%We construct four primary feature configurations: \textbf{RV-only}, \textbf{LV-only}, \textbf{MYO-only} and \textbf{ALL-structures} (concatenating RV, MYO, and LV features). To assess whether simple dynamic summaries add additional value, we also form explicit inter-phase delta and ratio features from the two labeled cardiac phases.

%\subsection{Evaluation Protocol}

%We evaluate models using balanced accuracy, macro-F1, and accuracy under 5-fold stratified cross-validation. To study label efficiency, we repeat training under multiple label fractions and report mean performance across repeated subsamples. To test whether observed results reflect genuine anatomical signal rather than artifacts, we include a label-shuffle sanity check in which pathology labels are randomly permuted. We report mean and standard deviation over 5-fold stratified cross-validation for the main ablation and model-comparison results. For label-fraction experiments, we additionally repeat random subsampling at each fraction and report the mean and standard deviation across repeated runs. All preprocessing steps, including median imputation and feature standardization when required, are fit within each training fold and then applied to the corresponding validation fold.

\subsection{Evaluation Protocol}

We evaluated models using accuracy, balanced accuracy and macro-F1 under 5-fold stratified cross-validation following standard multiclass classification metrics~\citep{sokolova2009systematic,powers2020evaluation}; refer \autoref{sec_math_formula_eval} for their expression. All preprocessing steps, including median imputation and feature standardization when required, are fit within each training fold and then applied to the corresponding validation fold. For label-fraction experiments, we additionally repeat random subsampling at each fraction and report mean and standard deviation across repeated runs.

%\section{Experiments}

%\subsection{Models}

%We compare three model families: \emph{multinomial logistic regression} \citep{hosmer2013applied}, RBF-SVM \citep{cortes1995support} and random forest \citep{breiman2001random}.
%\begin{itemize}
%    \item multinomial logistic regression,
%    \item RBF-SVM,
%    \item random forest.
%\end{itemize}
%These models span linear, kernel-based, and tree-based prediction while remaining lightweight and interpretable enough for a controlled representation study. Our aim is not to introduce a new method, but to determine whether classifier sophistication materially changes conclusions once the anatomical representation is fixed.

\subsection{Models}

We compare three lightweight model families: multinomial logistic regression as a linear baseline~\citep{hosmer2013applied,hastie2009elements}, RBF-SVM as a nonlinear kernel baseline~\citep{cortes1995support}, and random forest as a tree-based nonlinear baseline~\citep{breiman2001random}. %These models span linear, kernel-based, and ensemble decision boundaries while remaining lightweight enough for a controlled representation study. %We do not introduce a new predictor, but to determine whether classifier sophistication materially changes conclusions once the anatomical representation is fixed.
%For logistic regression, we used a multinomial formulation trained on standardized features. For SVM, we use an RBF kernel with standardized features and one-vs-rest multiclass decision functions. For random forest, we used an ensemble of decision trees trained directly on the imputed feature vectors without feature scaling. Across all models, missing feature values are imputed by the training-set median within each fold.
%Unless otherwise stated, the main comparisons use the following representative settings: multinomial logistic regression with default $\ell_2$ regularization, RBF-SVM with \texttt{C=10} and \texttt{gamma=scale}, and random forest with 300 trees. These choices were selected to provide competitive but lightweight baselines rather than heavily tuned model-specific optima. This keeps the comparison focused on representation quality rather than hyperparameter search. 
Additional details of the model that we considered are given in \autoref{tab:model_hparams}, \autoref{model_hyperparam}. Beyond aggregate predictive performance, we evaluated label efficiency, anatomy ablation, dynamic-feature enhancement, sanity checks, robustness, and feature-level interpretability.
%\subsection{Additional Analysis}
%We include several targeted analyses beyond aggregate performance that are essential to evaluate data efficiency, anatomical signal localization, robustness, and interpretability. \autoref{tab:analysis_overview} (\autoref{sec_additiona_details_tables-1}) summarizes the purpose of each component. %These analyses help distinguish genuine structure-dependent learning from accidental or model-specific effects.

\section{Results}
Our results address four complementary questions: whether the benchmark remains informative under limited labels, which anatomical structure carries the strongest predictive signal, whether simple inter-phase dynamic summaries add value beyond static anatomy-aware features, and whether the observed gains survive basic sanity checks. %Together, these analyses show that the benchmark is meaningful in the low-data regime and that its main signal is driven by anatomy-aware representation rather than by accidental shortcuts or simple feature augmentation.

\subsection{Label Efficiency}

\begin{figure}[h]
    \centering
    \includegraphics[width=.5\linewidth]{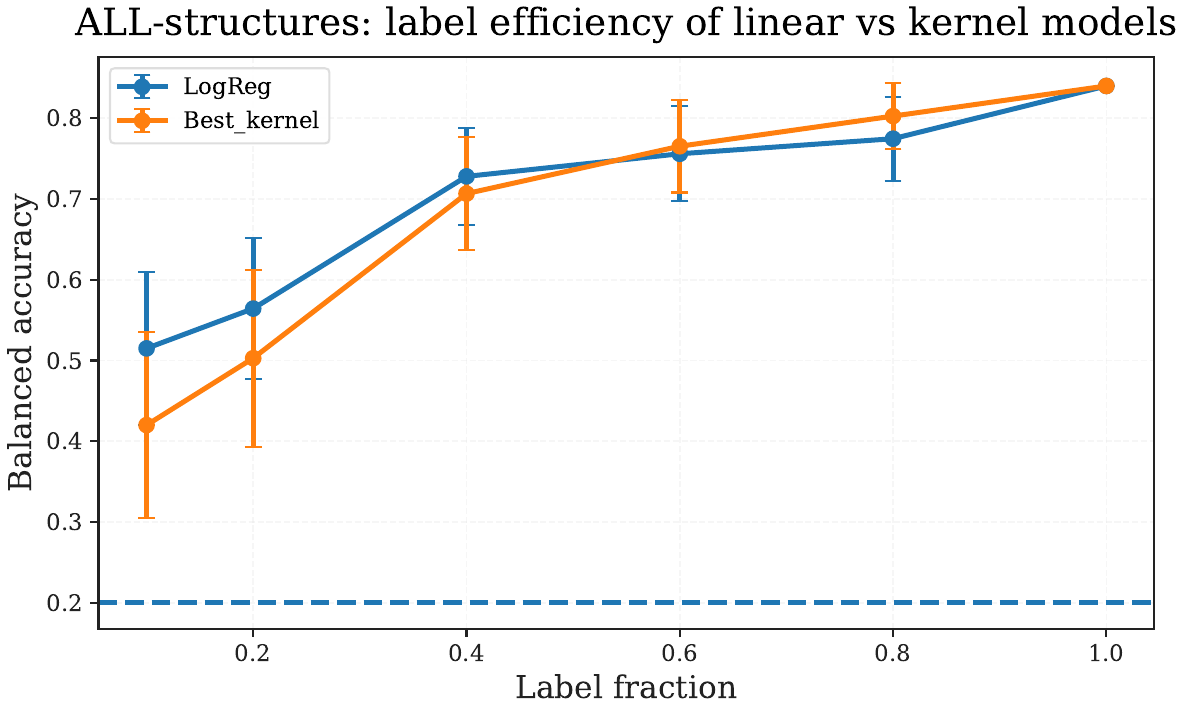}
    \caption{\textbf{Balanced accuracy across label fractions} for 5-class ACDC pathology prediction using the all-structure representation. %Performance improves steadily with additional labels, while linear and kernel models remain broadly similar across the low-data regime, indicating that anatomical representation contributes more than classifier complexity.
    }
    \label{fig:fig_1_label-fraction}
\end{figure}

We begin by evaluating whether segmentation-derived anatomical descriptors carry a meaningful pathological signal in the low-data regime. We observe that across repeated label-fraction sweeps, performance remains above chance ($1/5=0.2$, shown by the \textcolor{blue}{blue dashed line} in \autoref{fig:fig_1_label-fraction}) and improves gradually as additional labels are incorporated. Hence, this shows that the benchmark is neither trivial nor noise-dominated, and that simple patient-level anatomical features already support nontrivial multiclass prediction. A lightweight end-to-end ResNet-18 \citep{he2016deep} baseline trained on representative raw MRI slices performed substantially worse than all three anatomy-aware baselines (\autoref{sub_sec_end2endimage}-\autoref{tab:resnet_baseline}), reinforcing the value of explicit anatomical representation in the low-label regime.

%\autoref{fig:fig_1_label-fraction} establishes that the benchmark remains informative under limited supervision. Here, the performance increases steadily with additional labels being incorporated; however, the gap between linear and kernel models remains modest relative to the overall gain from having the all-structure representation itself. This supports the central claim of the paper that representation choice contributes more than classifier sophistication in the low-label regime.

\subsection{Which Anatomy Matters?}

%The central result of the paper is the anatomy ablation study. Among single-structure feature sets, MYO-only performs best, substantially outperforming LV-only and RV-only representations. The best overall performance is achieved by concatenating all structures. This indicates that pathology signal is not uniformly distributed across anatomy and that myocardial morphology carries the strongest single-structure information in this benchmark.
%The central result of the paper is the anatomy ablation study. Among single-structure feature sets, MYO-only performs best, substantially outperforming LV-only and RV-only representations. The best overall performance is achieved by concatenating all structures. Quantitatively, this indicates that pathology signal is not uniformly distributed across anatomy and that myocardial morphology carries the strongest single-structure information in this benchmark.

%This finding is further supported by qualitative visualizations: patient embeddings exhibit class structure in the full anatomical feature space, aligned class prototypes show visibly different myocardial contours across classes, and misclassifications concentrate in morphologically ambiguous cases rather than appearing random. 
%The anatomy ablation study provides the central empirical result of the paper. 

Now, we analyze the anatomy ablation study where among single-structure feature sets, MYO-only performs best, substantially outperforming both LV-only and RV-only representations, while the full multi-structure representation performs best overall. This suggests that the pathology signal is not distributed uniformly throughout the anatomy. Instead, myocardial morphology appears to concentrate the strongest single-structure information in this benchmark.

\begin{figure}[h]
    \centering
    \includegraphics[width=.5\linewidth]{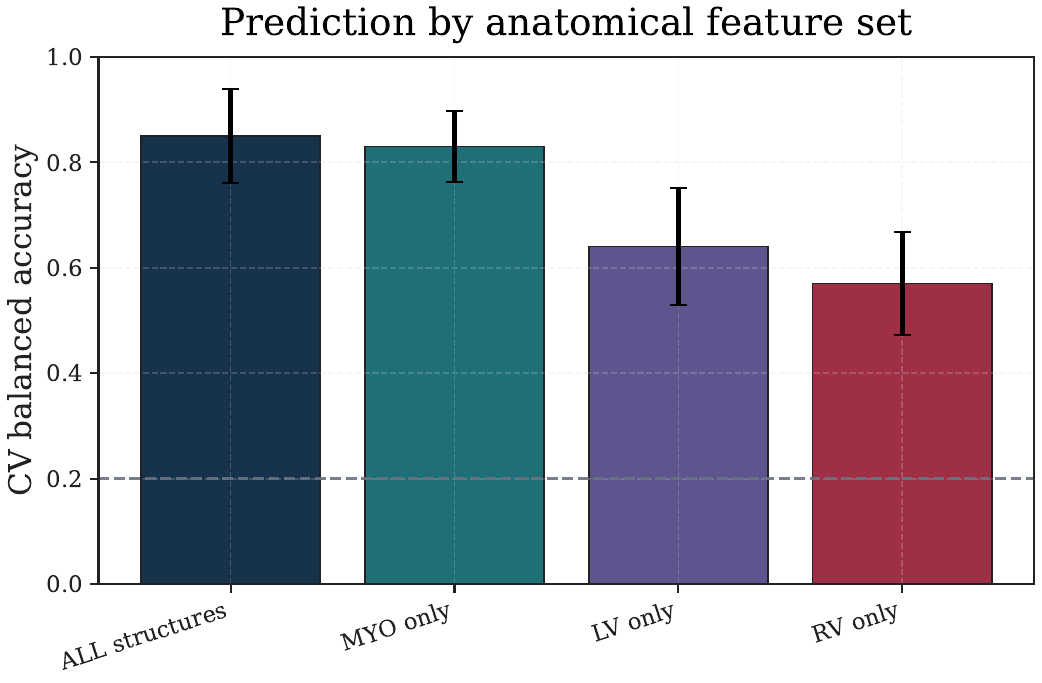}
    \caption{\textbf{Cross-validated balanced accuracy across anatomical feature} sets for 5-class ACDC pathology prediction. Myocardium is the strongest single-structure representation, while combining RV, myocardium, and LV yields the best overall performance.}
    \label{fig:fig_2_anatomy_abalation}
\end{figure}

From \autoref{fig:fig_2_anatomy_abalation}, we see that the myocardial descriptors are the strongest single-structure feature set, while the combined multi-structure representation performs best overall. In particular, the gain from moving from RV-only to MYO-only is much larger than the gain from switching among linear, kernel, and tree-based classifiers once the representation is fixed. %In other words, choosing the right anatomical structure matters more than choosing a more expressive predictive model. 
A finer-grained decomposition of feature importance by anatomical structure and descriptor family is provided in \autoref{fig:family_structure_heatmap}, \autoref{appendix_D}, further supporting the dominant role of myocardial descriptors.

We highlight that this result is important for exactly two reasons. First, on scientific grounds, our result suggests that the benchmark is driven less by generic whole-heart geometry and more by structure-specific morphology, with the myocardium serving as the most informative individual component. Second and lastly, from an empirical ground, our result provides an informative lesson for low-resource medical ML: when building simplified or compute-efficient pipelines under limited labels, the myocardium may be the most valuable single anatomical target to prioritize. All-structure representation still performs best, but the MYO result identifies where much of the predictive signal is already concentrated.

%The gain from moving from RV-only to MYO-only is substantially larger than the gain from switching among linear, kernel, and tree-based classifiers once the representation is fixed, indicating that anatomy-aware representation matters more than model sophistication in this benchmark.

%\subsection{Does Classifier Complexity Matter?}

%We next compare linear, kernel, and tree-based models on the strongest feature configuration. Although the best-performing nonlinear models slightly improve some classwise results, all model families lie in a relatively narrow performance band once the anatomical representation is fixed. In other words, the major gain does not come from increasing classifier sophistication. Instead, it comes from selecting the right anatomical factorization of the input. This result suggests that in low-data structured medical learning, representation quality may dominate classifier nonlinearity.

\subsection{Do Explicit Dynamic Features Help?}
To assess whether simple cardiac phase dynamics add information beyond static anatomical representation, we augment the full feature set with explicit inter-phase delta and ratio descriptors. These additions do not materially improve over the static multi-structure representation. We address this negative result cautiously. 

One possibility is that ACDC pathology groups are already strongly reflected in static morphology, particularly in myocardial structure, so that simple phase-difference summaries contribute little additional information. Another possibility is that our handcrafted dynamic descriptors are too compressed to preserve the richer spatial deformation patterns present across phases. Thus, our result should not be read as evidence that dynamics are uninformative in general; rather, it shows that simple low-dimensional inter-phase summaries do not outperform already-strong anatomy-aware static representations in this benchmark.
%\subsection{Do Explicit Dynamic Features Help?}
%To assess whether simple cardiac phase dynamics add information beyond static anatomical representation, we augment the full feature set with explicit inter-phase delta and ratio descriptors. We find that these additions do not materially improve over the static multi-structure representation. This suggests that the dominant pathology signal in this benchmark is already captured by anatomical geometry, at least for the simple handcrafted dynamic summaries considered here.
%\paragraph{Sanity Check}

For sanity check, under random label permutation, the balanced accuracy drops from \(0.870 \pm 0.057\) to \(0.230 \pm 0.057\), which is close to chance for a balanced 5-class task. This supports the interpretation that the observed gains arise from a genuine anatomical signal rather than leakage or spurious shortcut cues in the dataset.

%Under random label permutation, performance collapses to chance. This confirms that the benchmark is not being solved by leakage, label imbalance artifacts, or accidental shortcuts. Instead, the observed gains arise from the genuine anatomical signal present in the segmentation-derived features.

\subsection{Why Does MYO Matter?}
\begin{figure}[h]
    \centering
    \includegraphics[width=.5\linewidth]{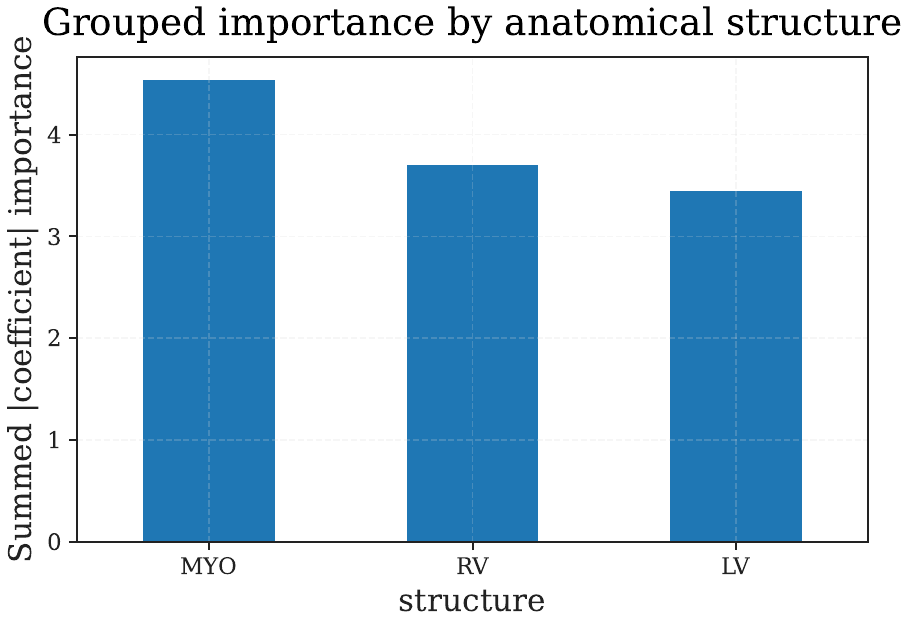}
\caption{\textbf{Grouped feature importance by anatomical structure.} Summed absolute logistic-regression coefficients are highest for myocardium, reinforcing the quantitative ablation result that MYO is the strongest single-structure source of predictive signal.}
    \label{fig:fig_8_group_imp_anatomical_struct}
\end{figure}

We provide a plausible explanation for the MYO result as follows: several ACDC pathologies are expressed strongly through the morphology of the myocardial wall rather than the geometry of the chamber geometry. Therefore, MYO descriptors capture clinically meaningful variation in shape, extent, circularity, elongation, and radial-distance structure, making the myocardium the most informative single anatomical compartment in the benchmark. %We aggregate feature importance by structure under the logistic-regression model to better understand the anatomical asymmetry observed in the ablation study. 
\autoref{fig:fig_8_group_imp_anatomical_struct} quantitatively helps explain the result of anatomy ablation by aggregating feature importance at the structure level. %Myocardium contributes the largest total importance, reinforcing the quantitative finding that MYO is the strongest single-structure source of signal. In this sense, the MYO result is not only an ablation outcome but also an interpretable feature-level conclusion.

\section{Discussion and Limitations}

Our key finding is that, in this low-data cardiac pathology benchmark, the dominant factor is the anatomical representation rather than the complexity of the classifier. %The strongest single-structure signal comes from the myocardium, and the best performance is obtained by combining myocardial and chamber geometry. 
Our reproducible benchmark experiments suggest that kernel and tree-based models provide limited gains beyond a strong anatomy-aware representation, while explicit handcrafted dynamic summaries add little further improvement. We note that our study also has limitations as we consider a single public dataset, rely on handcrafted segmentation-derived descriptors rather than raw-image end-to-end learning, and study dynamics only through simple inter-phase summaries. Our future work can extend this framework to additional datasets, uncertainty-aware analysis, richer temporal descriptors, and external validation across institutions.

\section{Conclusion}
%In this paper, we introduced an anatomy-aware benchmark for low-data cardiac pathology prediction using public ACDC segmentation masks. 
We empirically show that across anatomy ablations, model comparisons, dynamic-feature tests, and sanity checks, {the representation choice matters more than classifier complexity}. %Myocardial morphology is the strongest single-structure source of signal, while combining RV, MYO, and LV yields the best overall performance. 
These results suggest that in low-label structured medical learning, identifying the right anatomical factor may be more important than choosing a more expressive classifier.

%\section*{Acknowledgements}

\section*{Acknowledgments}

The author acknowledges the kind effort of \textsc{Payal Ghosh} (PG) in making \autoref{fig:fig_hook}. Additionally, we also acknowledge PG for helpful discussion and suggestions that were incorporated in the paper.%The overlays, anatomical decomposition, and benchmark-specific framing in the figure were added by the authors for the purposes of this work.

\section*{Impact Statement}
Our work has a broader impact in the design of practical medical AI systems.

\textbf{Disease-dependent anatomy-aware modeling.}
%This work is motivated by a practical question in medical AI: when labels and compute are limited, is it better to scale the model or to represent clinically meaningful structure more carefully? Our results support the latter direction in this benchmark. By building on a public cardiac MRI dataset, we contribute a reproducible low-data benchmark for anatomy-aware prediction~\citep{bernard2018acdc,blagec2023benchmarks}. This is especially relevant to resource-constrained healthcare environments, where prior work has identified data preparation, annotation, infrastructure, and implementation barriers as major bottlenecks for imaging AI~\citep{willemink2020preparing,varoquaux2022machine,mollura2020artificial}. 
%Our empirical findings suggest that anatomy-aware modeling may need to be 
%disease-specific, but other conditions may shift the anatomical priority. 
More generally, different cardiac conditions can shift the anatomical priority, suggesting that future low-label benchmarks should identify and emphasize structures that carry the most clinically meaningful information for the task at hand. This perspective may be relevant for the Global South settings, where data analysis %, computation, and annotation 
capacity are limited and where anatomically focused representations may offer a more practical path toward deployable medical AI.
%For instance, anomalies such as \citep{attenhofer2007ebstein} and recent clinical reports from countries of the global south, including Mexico, India, and China, illustrate its continued relevance in diverse healthcare settings \citep{herrera2020clinical,kumar2021ebstein,wu2024strategy}. This suggests that future low-label benchmarks for such conditions may need to prioritize the anatomical representation on the right side rather than the myocardial structure alone. %At the same time, our study is limited to a single public dataset and does not yet evaluate automated segmentation or external deployment, so we do not claim that anatomy-aware descriptors are universally superior to modern deep-learning pipelines.

\textbf{Data-efficient medical AI.}
%A central message of this work is that, in low-label medical imaging, performance may depend more on representing clinically meaningful anatomy than on escalating model complexity. %In our experiments, lightweight anatomy-aware models built from segmentation-derived descriptors substantially outperform a small end-to-end ResNet-18 baseline under the same limited-label setting. 
Our central message suggests that practical gains in medical AI may come not only from larger architectures but also from choosing the right anatomical representation. Recent learning from radiology and medical imaging has emphasized that limited data, limited expert labeling, and limited annotation resources present challenges for real-world model development~\citep{candemir2021training,willemink2020preparing}.

%\textbf{Reproducibility and benchmark design.}
%By building on the public ACDC dataset, this work contributes to the broader role of benchmark-driven progress in medical imaging AI. At the same time, recent work has argued that clinically relevant benchmark coverage remains limited and often misaligned with the needs of medical professionals. In this sense, our contribution is not only predictive but also methodological: we provide a transparent low-data benchmark for studying representation, label efficiency, and interpretability in a controlled cardiac setting~\citep{bernard2018deep,blagec2023benchmark,varoquaux2022machine}.

\textbf{Resource-constrained healthcare settings.}
%The findings are especially relevant to healthcare environments in which labels, compute, and deployment infrastructure are limited. Prior work has shown that medical imaging AI is often bottlenecked by data curation, annotation, standardization, and infrastructure rather than model choice alone~\citep{willemink2020preparing,varoquaux2022machine}. 
%Our work is fundamentally related to resource-constrained healthcare systems. In particular, previous work on AI in low and middle income countries further notes that scarce radiology resources can impede the adoption of computationally intensive imaging pipelines. %, while broader analyses of AI for health-system strengthening emphasize that successful deployment depends on workflows, local infrastructure, and implementation conditions rather than algorithmic accuracy alone~\citep{mollura2020artificial,ciecierski2022artificial,ahmed2023systematic}. 
Our results support a complementary actionable principle in resource-constrained healthcare settings: instead of assuming that better performance requires heavier end-to-end models, it may be more effective to identify the anatomical structures that carry the clinically meaningful signal and represent them explicitly.

\bibliography{example_paper}
\bibliographystyle{icml2026}

%%%%%%%%%%%%%%%%%%%%%%%%%%%%%%%%%%%%%%%%%%%%%%%%%%%%%%%%%%%%%%%%%%%%%%%%%%%%%%%
%%%%%%%%%%%%%%%%%%%%%%%%%%%%%%%%%%%%%%%%%%%%%%%%%%%%%%%%%%%%%%%%%%%%%%%%%%%%%%%
% APPENDIX
%%%%%%%%%%%%%%%%%%%%%%%%%%%%%%%%%%%%%%%%%%%%%%%%%%%%%%%%%%%%%%%%%%%%%%%%%%%%%%%
%%%%%%%%%%%%%%%%%%%%%%%%%%%%%%%%%%%%%%%%%%%%%%%%%%%%%%%%%%%%%%%%%%%%%%%%%%%%%%%
\newpage
\appendix
\onecolumn

\section{Evaluation and Model Details}\label{sec_math_formula_eval}
\subsection{Metric Definitions}
Let \(C_{ij}\) denote the confusion matrix entry counting examples whose true class is \(i\) and predicted class is \(j\). For class \(i\), the recall is defined as follows: $\mathrm{Recall}_i = \frac{C_{ii}}{\sum_j C_{ij}}.$
Then, \emph{balanced accuracy} $\mathrm{BA}$ is defined as the average recall across classes: $\mathrm{BA} = \frac{1}{K}\sum_{i=1}^K \mathrm{Recall}_i,$ where \(K\) is the number of classes. This metric is appropriate here because the benchmark is multiclass and class-conditional performance is of primary interest. In the usual manner, \emph{accuracy} is defined as follows: $\mathrm{Acc} = \frac{\sum_i C_{ii}}{\sum_{i,j} C_{ij}}.$ For macro-F1, let the precision for class \(i\) be $\mathrm{Precision}_i = \frac{C_{ii}}{\sum_j C_{ji}},$
and define the classwise F1 score by $\mathrm{F1}_i = \frac{2\,\mathrm{Precision}_i\,\mathrm{Recall}_i}{\mathrm{Precision}_i+\mathrm{Recall}_i}.$
The macro-F1 score is then $\mathrm{Macro\text{-}F1} = \frac{1}{K}\sum_{i=1}^K \mathrm{F1}_i.$

%%%%%%%%%%%%%%%%%%%
% \input{table_why}
%%%%%%%%%%%%%%%%%%%

\subsection{Model Hyperparameters}\label{model_hyperparam}
Other important details related to the models that we consider in the paper.
\begin{table}[h]
\centering
\small
\caption{Model families and representative hyperparameter settings used in the main experiments. %These settings were chosen to provide competitive but lightweight baselines rather than aggressively tuned model-specific optima.
}
\label{tab:model_hparams}
\begin{tabular}{p{3cm}p{4cm}p{5.5cm}}
\toprule
\textbf{Model} & \textbf{Role in study} & \textbf{Main settings} \\
\midrule
Multinomial logistic regression
& Linear baseline for testing whether simple decision boundaries suffice once anatomical representation is fixed
& Multinomial formulation; median imputation; feature standardization; default \( \ell_2 \) regularization; trained with standard convex optimization \\
\addlinespace

RBF-SVM
& Nonlinear kernel baseline for testing whether modestly more expressive decision boundaries improve over the linear baseline
& RBF kernel; \(C=10\); \texttt{gamma=scale}; one-vs-rest multiclass decision function; median imputation; feature standardization \\
\addlinespace

Random forest
& Tree-based nonlinear baseline for testing whether ensemble partitioning improves over linear and kernel models
& 300 trees; median imputation; no feature standardization; default impurity-based splits and bagging-style ensemble aggregation \\
\bottomrule
\end{tabular}
\end{table}

\section{Additional Qualitative Analysis}
\begin{figure}[h]
    \centering
    \includegraphics[width=.4\linewidth]{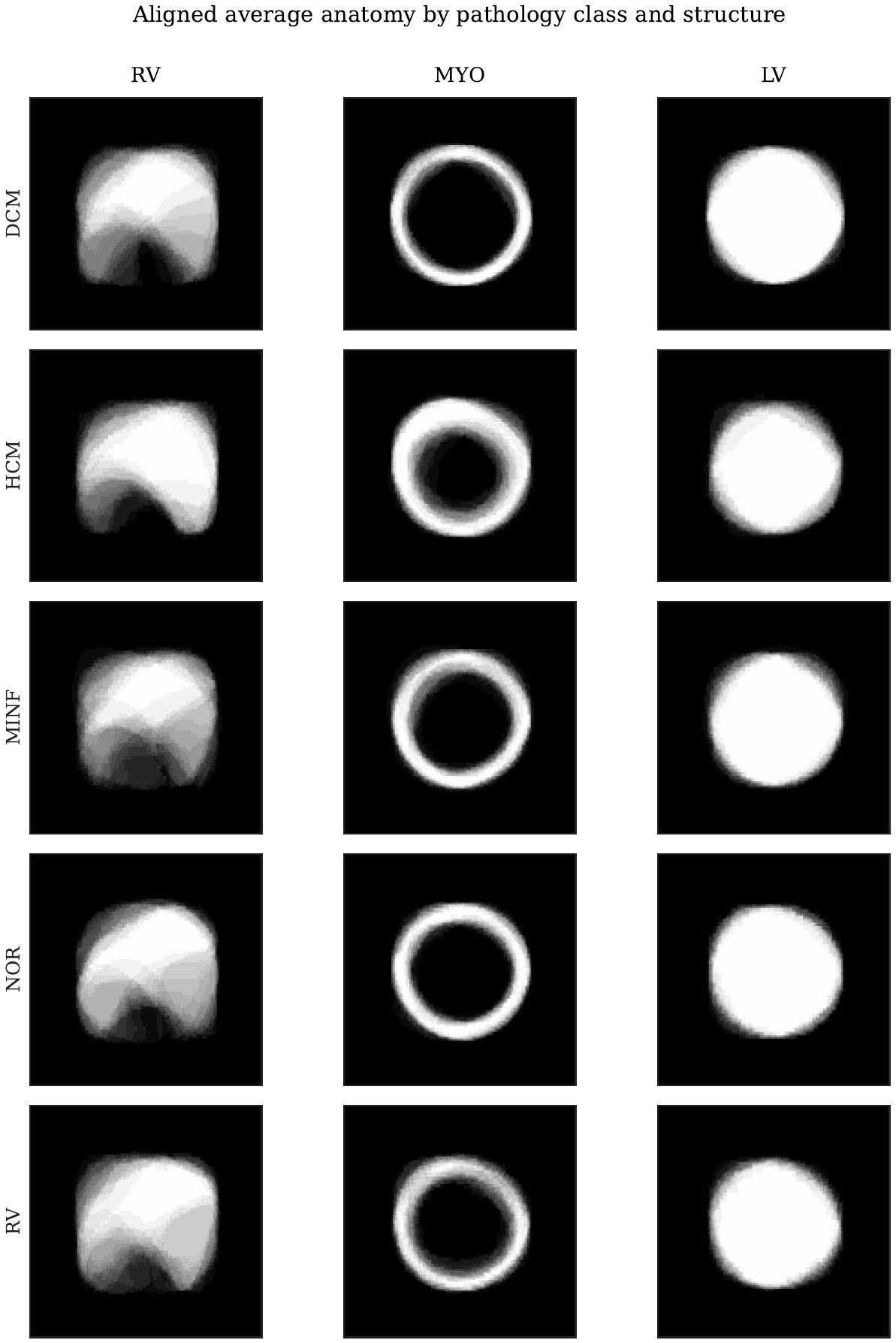}
\caption{Aligned class prototypes for RV, myocardium, and LV across ACDC pathologies. Masks were centered and size-normalized before averaging. Myocardial contours exhibit the clearest class-dependent variation, consistent with the quantitative finding that myocardium is the strongest single-structure feature set.}
    \label{fig:fig_6-1_aligned_avg_anatomy}
\end{figure}

To visually complement the anatomical ablation study, we align and average structure-specific masks within each pathology class and plot the resulting class prototypes; this is visually shown in \autoref{fig:fig_6-1_aligned_avg_anatomy}.

\subsection{Representative Failure Cases}

\begin{figure}[h]
    \centering
    \includegraphics[width=.45\linewidth]{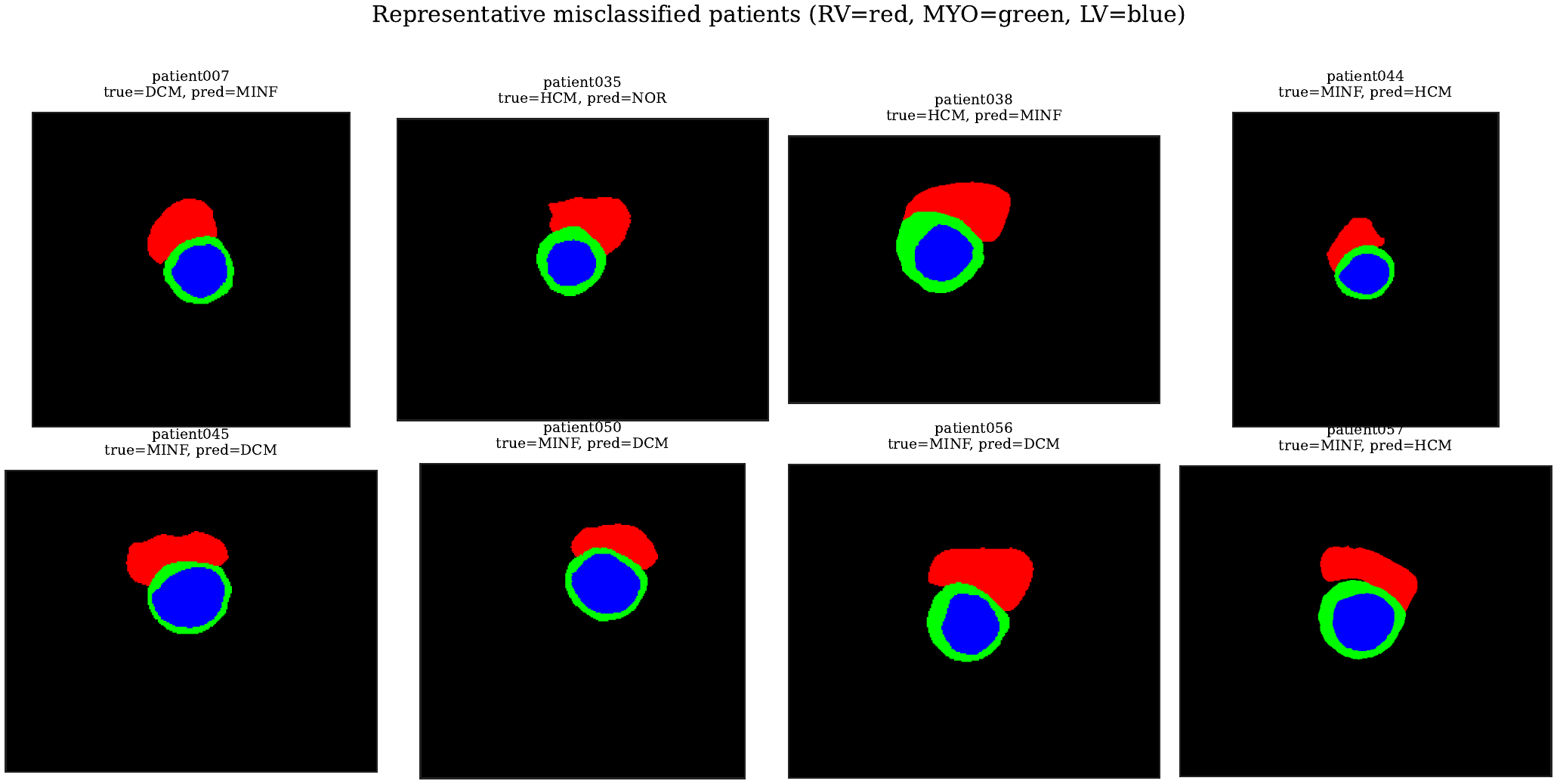}
\caption{Representative misclassified patients under the all-structure model. RV, myocardium, and LV are shown in red, green, and blue, respectively. The errors are concentrated in anatomically ambiguous cases rather than random failures, especially for MINF-related confusions.}
    \label{fig:fig5_rep_miss_patient}
\end{figure}

To complement the confusion analysis, we visualize representative misclassified patients to illustrate the kinds of anatomical ambiguity that remain under the all-structure representation; this is shown in \autoref{fig:fig5_rep_miss_patient}.

\section{Additional Quantitative Analysis}

\subsection{Normalized Confusion Matrix}

\begin{figure}[h]
    \centering
    \includegraphics[width=.25\linewidth]{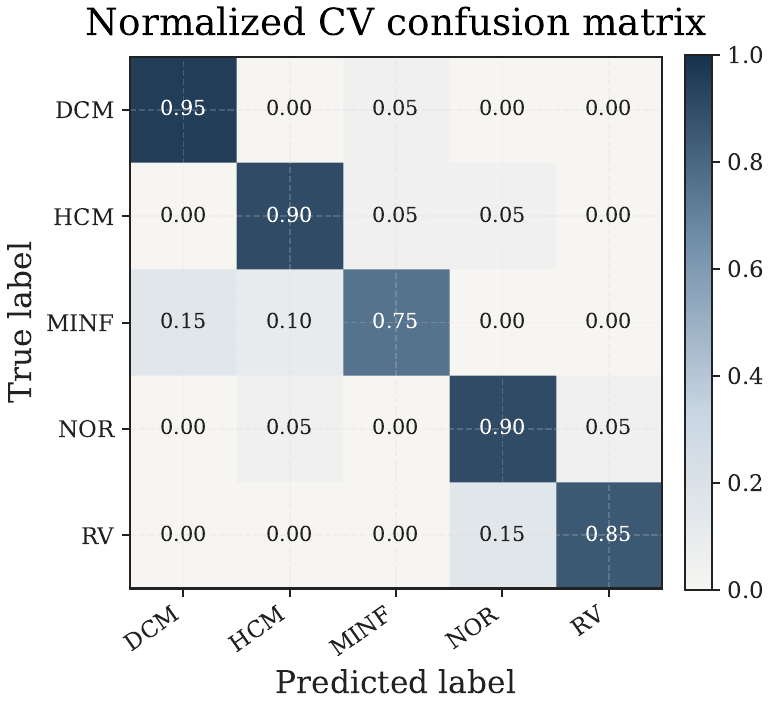}
    \caption{Normalized cross-validated confusion matrix for 5-class ACDC pathology prediction using the all-structure anatomical representation. Most classes are well separated, while residual errors are concentrated in a small number of structured class pairs, particularly NOR--RV and MINF--DCM, indicating a meaningful but nontrivial benchmark.}
    \label{fig:fig_4_aligned}
\end{figure}

To characterize residual class-wise ambiguity, we compute the row-normalized confusion matrix. Let \(C_{ij}\) denote the number of examples whose true class is \(i\) and predicted class is \(j\). We normalize each row by the total number of examples in the corresponding true class:
\begin{align*}
    \widetilde{C}_{ij} = \frac{C_{ij}}{\sum_{k} C_{ik}}.
\end{align*}
Thus, \(\widetilde{C}_{ij}\) represents the fraction of class-\(i\) examples assigned to class \(j\), and each row sums to one. \autoref{fig:fig_4_aligned} shows that the remaining errors are structured rather than random. DCM and RV are especially recovered well, while MINF and NOR account for most of the residual ambiguity. This pattern suggests that the benchmark is meaningful but not saturated, with failures concentrated in a small number of clinically plausible class pairs.

\subsection{Robustness to Imperfect Segmentations}
\begin{figure}[h]
    \centering
    \includegraphics[width=.35\linewidth]{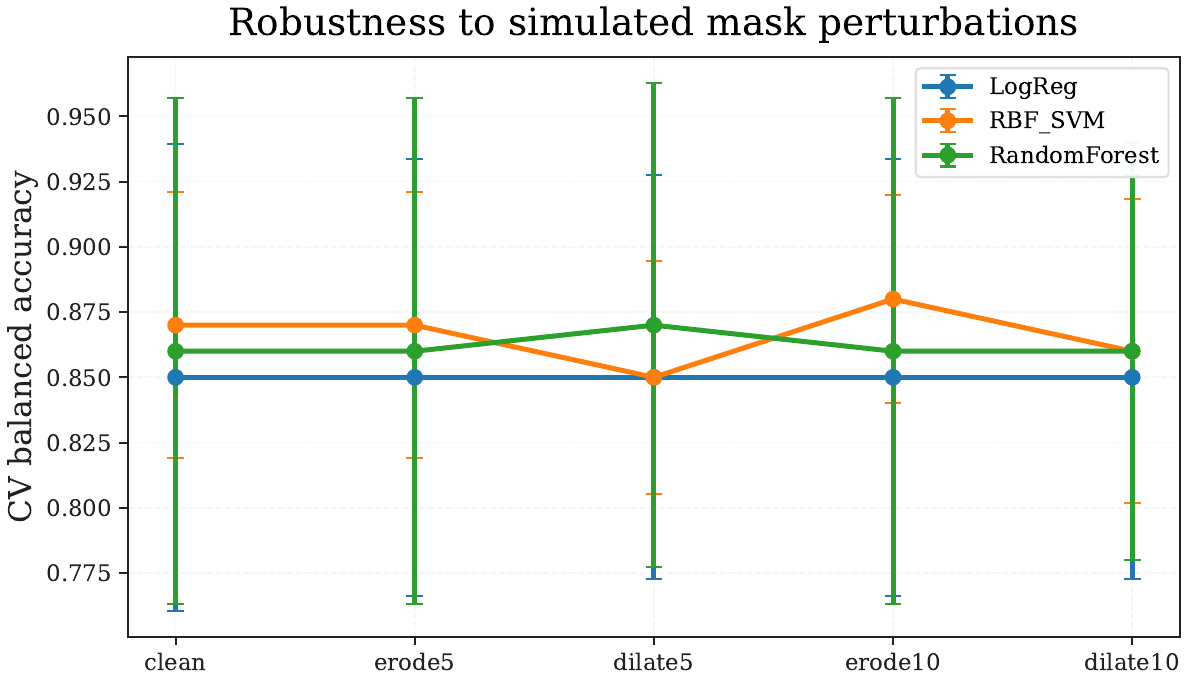}
\caption{Robustness to simulated mask perturbations. We erode and dilate segmentation masks by small amounts to mimic annotation disagreement or lower-quality imaging conditions. Performance remains relatively stable across mild perturbations, suggesting that the anatomy-aware pipeline is robust to realistic segmentation noise.}
    \label{fig:fig_7_robustness_mask_pert}
\end{figure}

To assess whether the anatomy-aware pipeline remains reliable under realistic segmentation imperfections, we simulate mild boundary perturbations through mask erosion and dilation and re-evaluate the benchmark under the same cross-validation protocol. \autoref{fig:fig_7_robustness_mask_pert} tests whether the pipeline remains reliable under imperfect segmentations. Performance remains relatively stable under mild erosion and dilation perturbations, suggesting that the anatomy-aware representation is not unduly brittle to realistic contour variability. This is particularly relevant for deployment in resource-constrained settings, where segmentation quality may vary due to annotation differences or heterogeneous imaging conditions.

\subsection{Feature-Level and Representation Analysis}

\begin{figure}[h]
    \centering
    \includegraphics[width=.3\linewidth]{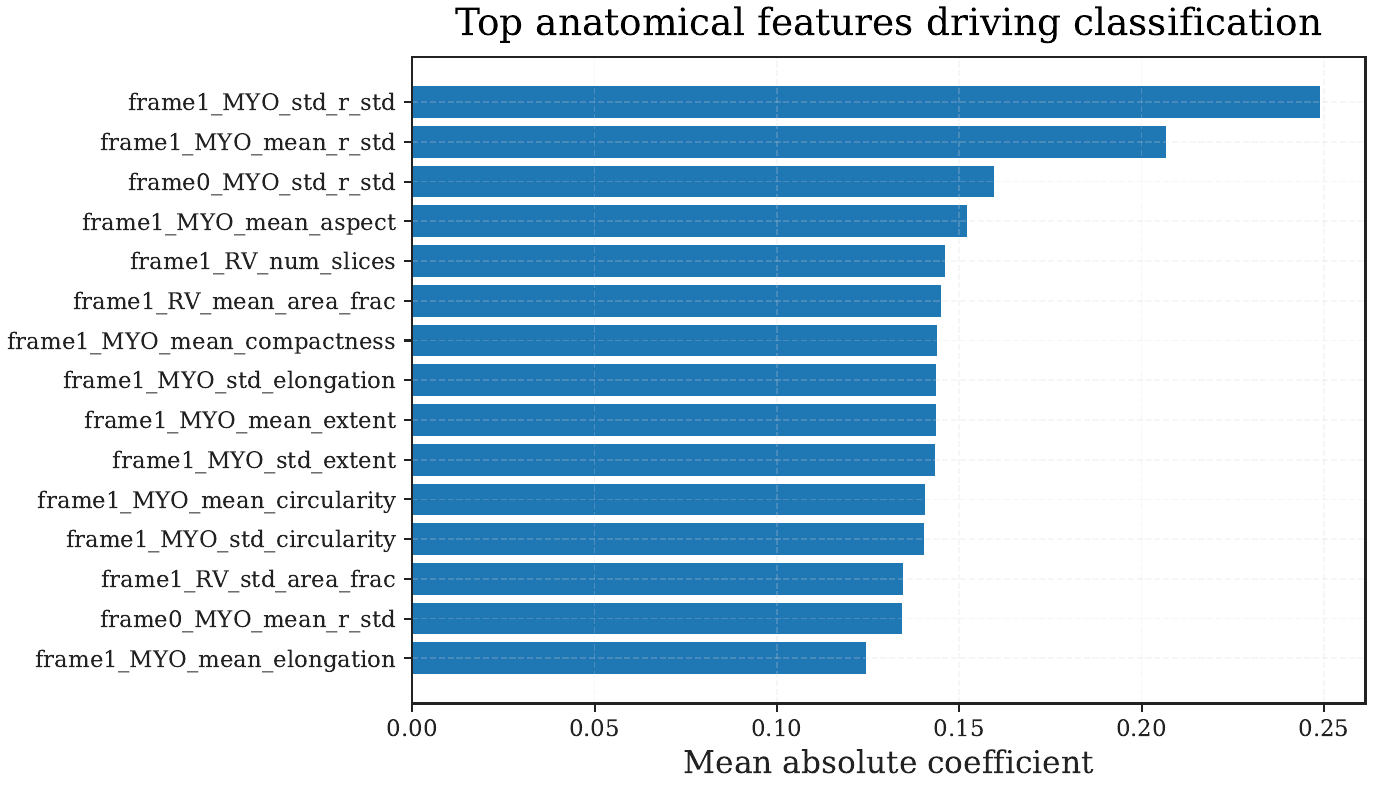}
\caption{Top anatomical features driving classification under the logistic-regression model, ranked by mean absolute coefficient magnitude. Many of the most influential descriptors arise from myocardial morphology, with additional contribution from a smaller set of RV-derived features.}
    \label{fig:fig_9_top_anatomical_struct}
\end{figure}

\autoref{fig:fig_9_top_anatomical_struct} refines the MYO story by showing which individual descriptors drive the classification. Many of the top-ranked features are myocardial in origin, with only a smaller contribution from RV-derived quantities. This suggests that the predictive value of MYO is not a coarse artifact of grouping, but is expressed through specific shape statistics extracted from the myocardium itself.

\subsection{Patient embedding}

\begin{figure}[h]
    \centering
    \includegraphics[width=.3\linewidth]{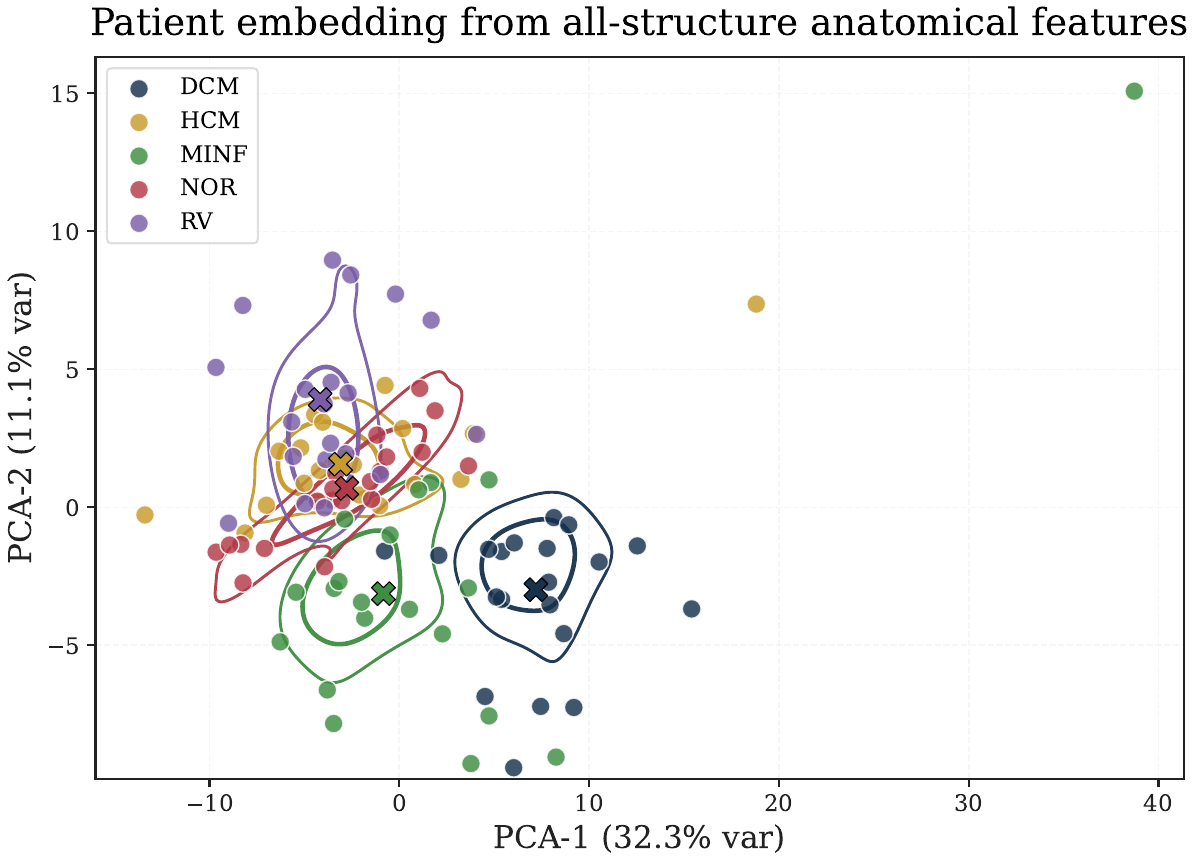}
    \caption{PCA embedding of patients in the all-structure anatomical feature space. The classes exhibit visible organization without becoming trivially separable, indicating that the anatomy-aware representation captures meaningful pathology structure while preserving nontrivial class overlap.}
    \label{fig:fig3_patient_embedding_refined}
\end{figure}

\autoref{fig:fig3_patient_embedding_refined} shows that the resulting patient-level feature space has visible class structure, supporting the interpretation that the representation captures clinically meaningful anatomical organization rather than random variation.

\subsection{Most important myocardial descriptor families}

\begin{figure}[h]
    \centering
    \includegraphics[width=.4\linewidth]{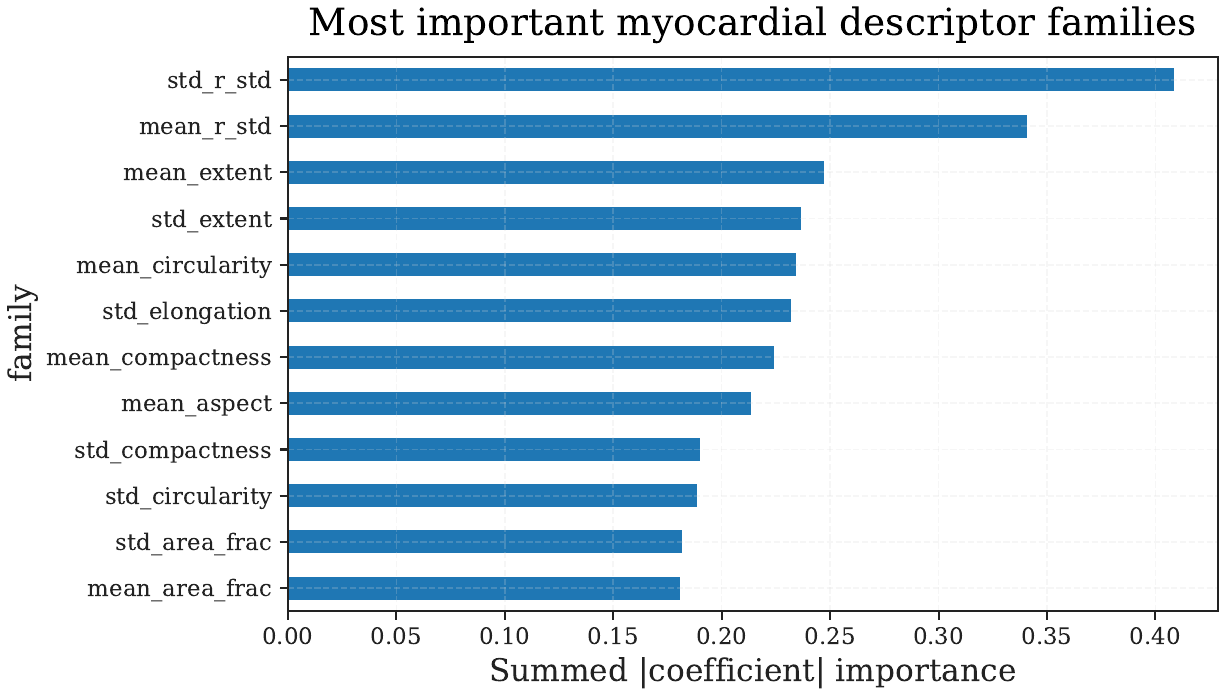}
\caption{Most important myocardial descriptor families under the logistic-regression model. Radial-distance variability, extent, circularity, elongation, and compactness emerge as the most influential myocardial descriptor groups, suggesting that the predictive value of MYO arises from geometry rather than a single scalar measurement alone.}
    \label{fig:fig_10_most_imp_myo_descrip}
\end{figure}

\autoref{fig:fig_10_most_imp_myo_descrip} further decomposes the myocardial contribution into descriptor families. Radial-distance variability, extent, circularity, elongation, and compactness dominate, indicating that the MYO signal arises from geometry and morphological variation rather than from a single scalar quantity. This deepens the interpretability of the benchmark and suggests which types of myocardial structure are most informative with limited labels.

\section{Structure-by-Descriptor-Family Importance}\label{appendix_D}
\begin{figure}[h]
    \centering
    \includegraphics[width=.4\linewidth]{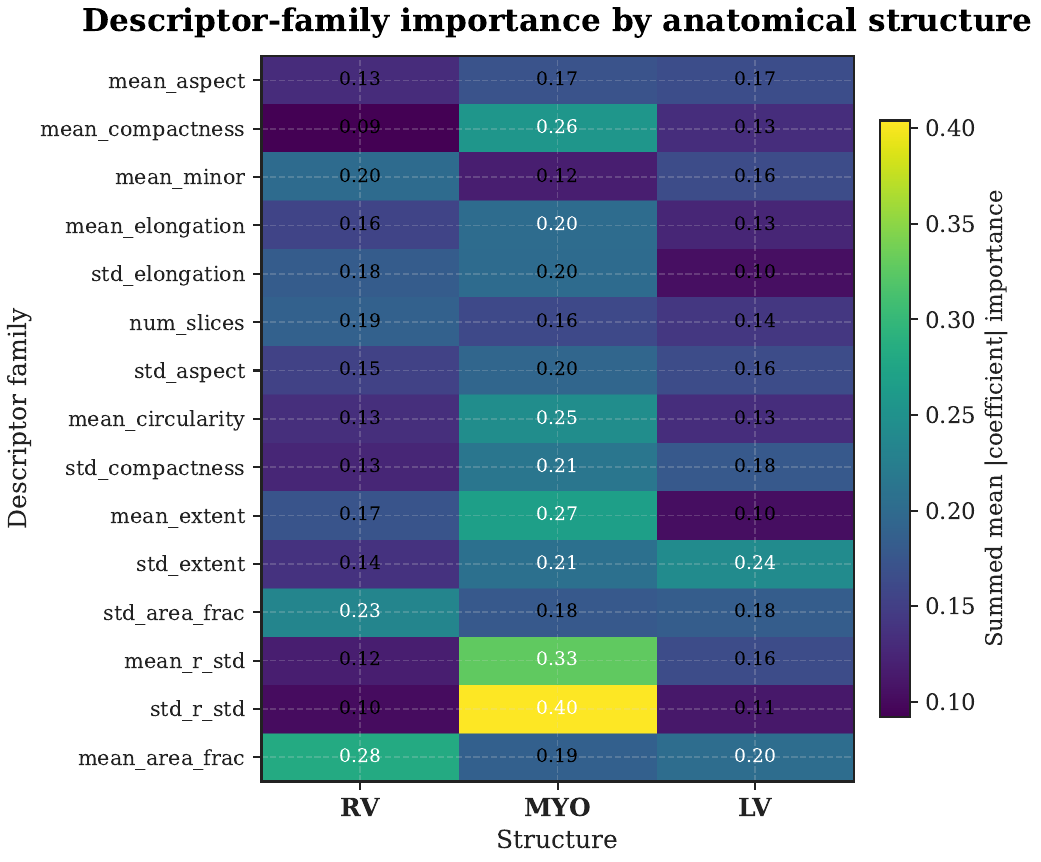}
\caption{Structure-by-descriptor-family importance analysis using multinomial logistic regression coefficients. Each cell reports the summed mean absolute coefficient importance for a descriptor family within a given anatomical structure, averaged across cross-validation folds. }
    \label{fig:family_structure_heatmap}
\end{figure}

In \autoref{fig:family_structure_heatmap}, the heatmap provides a finer-grained interpretability view of the anatomy-aware benchmark and further highlights the dominant contribution of myocardial descriptors.
\subsection{Robustness}
\begin{figure}[h]
    \centering
    \includegraphics[width=0.5\linewidth]{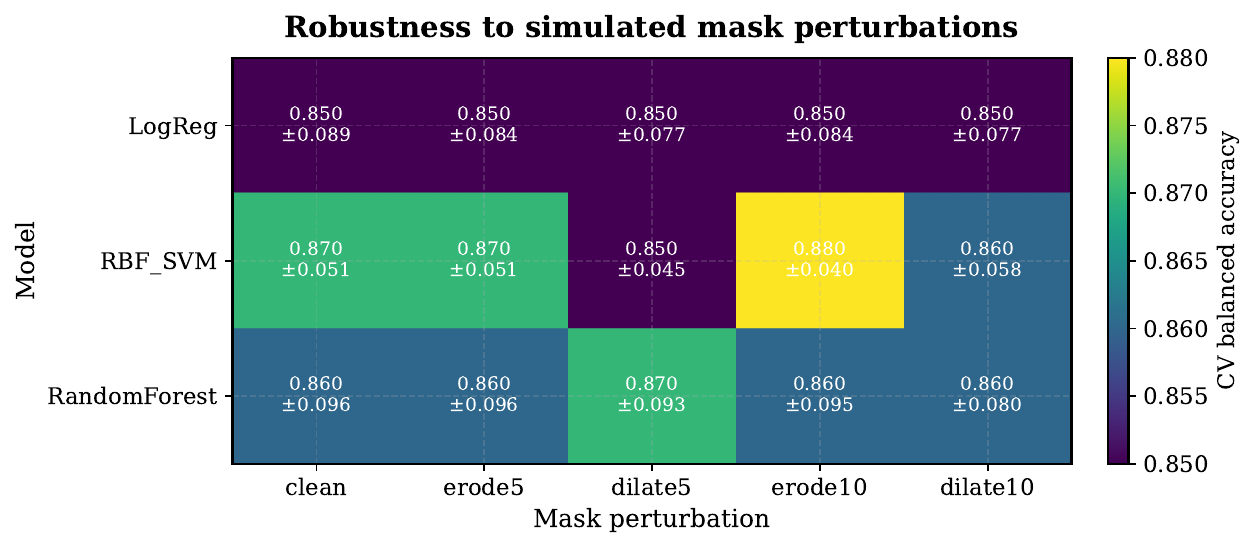}
\caption{Robustness to simulated mask perturbations. Each cell reports mean cross-validation balanced accuracy \(\pm\) standard deviation under mild erosion and dilation of the segmentation masks. Performance remains stable across perturbation settings for logistic regression, RBF-SVM, and random forest, suggesting that the anatomy-aware pipeline is robust to modest contour variability.}
    \label{fig:robustness_heatmap}
\end{figure}

\section{End-to-End Image Baseline}\label{sub_sec_end2endimage}
To contextualize anatomy-aware baselines, we additionally compared with a lightweight end-to-end image classifier based on ResNet-18 (parameter counts 11,175,941) trained on representative raw MRI slices under the same low-label protocol.
\begin{table}[h]
    \centering
    \caption{Comparison of lightweight anatomy-aware baselines and a small end-to-end image baseline (ResNet-18). The anatomy-aware models operate on segmentation-derived patient descriptors, whereas ResNet-18 operates on representative raw MRI slices. This comparison contextualizes the tradeoff between predictive performance and computational cost in low-resource settings.}
    \begin{tabular}{lllllll}
\toprule
Model & Input Type & Balanced Acc. & Macro-F1 & Train time/fold (s) & Infer time/fold (s) \\
\midrule
LogReg & anatomy features & 0.85 ± 0.089 & 0.849 ± 0.089 & - & - \\
RBF-SVM & anatomy features & 0.87 ± 0.051 & 0.870 ± 0.051 & - & - \\
RandomForest & anatomy features & 0.86 ± 0.097 & 0.853 ± 0.102 & - & - \\
ResNet18 (2-channel) & raw MRI slices & 0.41 ± 0.074 & 0.380 ± 0.083 & 5.45 & 0.02 \\
\bottomrule
\end{tabular}
    \label{tab:resnet_baseline}
\end{table}

%%%%%%%%%%%%%%%%%%%%%%%%%%%%%%%%%%%%%%%%%%%%%%%%%%%%%%%%%%%%%%%%%%%%%%%%%%%%%%%
%%%%%%%%%%%%%%%%%%%%%%%%%%%%%%%%%%%%%%%%%%%%%%%%%%%%%%%%%%%%%%%%%%%%%%%%%%%%%%%

\end{document}